\documentclass[journal]{IEEEtran}
\ifCLASSINFOpdf
 \usepackage[pdftex]{graphicx}
\else
\fi
\usepackage{lipsum}% http://ctan.org/pkg/lipsum
\usepackage{algorithm2e}
\usepackage{epstopdf}

\usepackage{cite}
\usepackage{booktabs}

\usepackage[caption=false]{subfig}
\usepackage{amsmath}
\usepackage{amsfonts}

\begin{document}
\title{A Stochastic Model for Uncontrolled Charging of Electric Vehicles Using Cluster Analysis}

\author{Constance~Crozier,~\IEEEmembership{Student~Member,~IEEE,}
        Thomas~Morstyn,~\IEEEmembership{Member,~IEEE}
        and~Malcolm~McCulloch,~\IEEEmembership{Senior~Member,~IEEE}
\thanks{C. Crozier, T. Morstyn, and M. McCulloch are with the Department of Engineering Science, University of Oxford.}}% <-this % stops a space

% note the % following the last \IEEEmembership and also \thanks -- 

\markboth{Journal of \LaTeX\ Class Files,~Vol.~14, No.~8, August~2015}%
{Shell \MakeLowercase{\textit{et al.}}: Bare Demo of IEEEtran.cls for IEEE Journals}



\maketitle
\begin{abstract}
This paper proposes a probabilistic model for uncontrolled charging of electric vehicles (EVs). EV charging will add significant load to power systems in the coming years and, due to the convenience of charging at home, this is likely to occur in residential distribution systems. Estimating the size and shape of the load will allow necessary reinforcements to be identified. Models predicting EV charging are usually based on data from travel surveys, or from small trials. Travel surveys are recorded by hand and typically describe conventional vehicles, but represent a much larger and more diverse sample of the population. The model here utilizes both sources: trial data to parameterize the model, and survey data as the model input. Clustering is used to identify modes of vehicle use, thus reducing vehicle use to a single parameter -- which can be incorporated into the model without adding significant computational burden. Two case studies are included: one investigating the aggregated charging of 50 vehicles, and one predicting the increase in after diversity maximum demand for different regions of the UK. 
\end{abstract}


% \fi
\IEEEpeerreviewmaketitle

\section{Introduction}
\IEEEPARstart{T}{his} paper proposes a stochastic model for the charging power demand of electric vehicles (EVs) -- designed to be used in future planning for transmission and distribution systems. The model combines high temporal-resolution data from a small-scale EV trial data, with national travel survey data which primarily describes conventional vehicles. 

\par{EVs represent a rapidly increasing share of the vehicle fleet; it is forecast that there could be 36 million EVs on UK roads by 2040~\cite{FES18}. This will contribute significantly to the \(\text{CO}_{\text{2}}\) emissions reduction required to meet the Paris Climate agreement~\cite{paris}. However, charging of EVs will present challenges for the power system; peak demand~\cite{Wu2011}, system losses~\cite{Leou2014}, and voltage violations~\cite{Dubey2015} are all expected to rise as a result of the additional load. Smart charging of vehicles could protect the network without affecting mobility~\cite{Xiong2015}. However, while trials are being carried out, there are currently no incentives for consumers to alter their charging so as to protect the grid. Therefore, it is important that consumers' behaviour without smart charging can be accurately modelled, in order to identify required system upgrades.}

% stochastic
\par{Power systems are designed to operate under uncertain loading, with some degree of confidence. Therefore, when planning for the future, it is insufficient to estimate the average load due to EV charging -- the variability also needs to be considered. Stochastic load models output a probability distribution of power demand rather than a single estimate. In the case of EV charging, there are two sources of variability: the vehicle use, and the charging behaviour. The first describes variations in travel behaviour, both between users and day-to-day. The second describes variations in the circumstances under which a user will charge their vehicle. These must both be modelled in order to fully capture the variability in charging. Stochastic models for EV charging can be broadly decomposed into three groups: deterministic models applied to stochastic vehicle use~\cite{Wu2011,Huang2010,Darabi2011,Yan2017,Leou2014,Barghi-Nia2015,Pashajavid2012,Klayklueng2015,Ahmadian2015,Hilton2018}, stochastic models applied to deterministic vehicle use~\cite{Shahidinejad2012,Omran2014}, and top down stochastic charging models~\cite{Godde2015,Quiros-Tortos2018,Rolink2013,6595730,Liang2014}.}

\par{The first group encompasses the majority of the early research in this area. In these models simple assumptions are made for charging -- e.g. that it begins after completion of the final journey of the day, or anytime the vehicle is home. Variation in predicted charging is then due only to varied vehicle use, which is captured by sampling either raw vehicle data (e.g. \cite{Wu2011}), or probability distribution functions (PDFs) for energy use and arrival times (e.g. \cite{Huang2010,Darabi2011,Yan2017,Leou2014,Barghi-Nia2015,Pashajavid2012,Klayklueng2015,Ahmadian2015,Hilton2018}). Providing the data source is large and representative, these models will capture variability in vehicle use. However, they do not include variability introduced by users' charging decisions -- all variability will be due to the distribution of arrival times. Diversity generally results in lower aggregated loads, so these models are likely overestimating peak demand.}

\par{The second group of models take a given vehicle use, and produce a stochastic estimate of charging. Creating these generally requires data where both the use and charging of EVs are recorded. Fuzzy logic models are used in \cite{Shahidinejad2012,Omran2014}, where certain combinations of input parameters result in a low, medium or high probability of charging. In \cite{Shahidinejad2012} the vehicles' state-of-charge (SOC) and length of parking time are assumed to impact the users' decision to charge, while \cite{Omran2014} also incorporates the distance from home. Considering only 3 probability states limits the accuracy of these models, however further states introduce additional parameters which require a large amount of data to set confidently.}

\par{The third group directly models charging, rather than the relationship between vehicle use and charging. In other words, these are top-down models for EV charging. Sometimes standard probabilistic models are used: Gaussian Mixture Models are used in \cite{Godde2015,Quiros-Tortos2018}, and \cite{Rolink2013} uses a non-homogenous Markov Process. In \cite{6595730,Liang2014} random point processes are used to describe EV arrivals, and queueing theory is used to model EV charging. However, this approach is perhaps better suited to public charging, where the availability of the charger is a limiting factor. These models likely capture the variability from their constituent datasets, but also any sources of bias present in the data. Also, as they do not accept inputs, they can not be applied to a different set of vehicle usage.}

\par{All stochastic charging models require data, and those in the first group predominantly use travel surveys. These datasets are typically large, and contain regional information -- allowing geographic variation to be considered. However, they primarily describe conventional vehicles, so no charging behaviour is recorded, and the accuracy of the data is limited by human error. The second two groups require charging data, so are typically based on data from small scale EV trials. As these trials are opt-in, the participants are likely to be a biased subset of the UK's drivers. In \cite{Haustein2018} it is suggested that early EV adopters are likely to have high incomes and more than one vehicle, which would result in a narrower set of vehicle use. Therefore, extrapolating these trials' data to represent a larger fleet of vehicles' charging is unlikely to produce accurate results. Additionally, these trials are small and in sparse geographic locations, so the regional variation in EV charging can not be investigated. To the authors' knowledge, no previously proposed models have combined both sources of data.}

%\par{It is likely that the charging behaviour of EVs will vary regionally -- as vehicle usage varies significantly across the country. Currently, EV usage data is only available from small scale trials taking part in specific areas.  Travel surveys are routinely carried out all over the world, aiming to provide insight into the way transport is used by the population. Although this data primarily represents conventional vehicle usage, its size and geographic breadth allows local trends to be investigated. Therefore, a model which can be applied to this kind of data allows for more in depth analysis.}

\par{Vehicle usage data, such as that recorded in travel surveys, is high dimensional -- as the timings and distance of a potentially large number of journeys are recorded. Clustering allows data to be grouped, thereby reducing the dimension to a single parameter. The clustering in this paper extends the method first presented by the authors in \cite{Crozier2018ISGT}, by considering types of driving days instead of types of driver. Clustering of vehicle trajectories is a more mature research topic  (e.g. \cite{5462900}), however the aim of these works is to identify common origin-destination pairs. Here we consider the broader problem of identifying days of vehicle use which are temporally similar.}

\par{In this paper we present a stochastic model for EV charging which is parameterized by trial data, but can be applied to survey data -- thus combining the benefits of both sources of data. The success of such a model can be quantified how accurately it models charging from the trial data. However, applying it to the survey data provides further insight, particularly on the likely regional variation in charging behaviour. The model uses clustering to identify different vehicle use cases, and considers the variation in charging behaviour they exhibit. Therefore, the relationship between vehicle use and charging is incorporated into the model without introducing a large number of parameters. As the charging model can be repeated using different vehicle data, variation in both travel behaviour and user charging can be considered.}
 
 \par{The contributions of this paper can be summarised as follows: First, that we use clustering to identify typical modes of vehicle use in the UK. Second, that we quantify the differences between vehicle usage in early EV trials, and that exhibited in the National Travel Survey (which is representative of the UK fleet as a whole). Third, that we formulate a stochastic model for charging which combines travel survey and trial data -- allowing both uncertainty in charging and vehicle use to be modelled simultaneously. Finally, that we set up realistic case studies for EV charging demand in the UK, predicting the regional variation of the impact of EV charging on distribution networks.}

% scope
\par{Only at-home charging is considered here; further work and data would be required to adapt this model for public charging use. The model parameters and results presented in this paper are specific to the UK. However, the methodology could be applied to equivalent data from other countries.}

% structure
\par{The remainder of this paper is structured as follows: In Section \ref{sec:data} the data used in this analysis are described, Section \ref{sec:clst} contains the clustering methodology and analysis, the proposed model is described in Section \ref{sec:cm} and validated in Section \ref{sec:val}, case studies are presented in Section \ref{sec:res}, and Section \ref{sec:con} concludes the paper.}

\section{Vehicle Usage Data}\label{sec:data}

\par{The proposed model combines both survey and trial data, the specific data sources used are described in this section.}

\subsection{Travel Survey Data}

\par{Many countries carry out travel surveys, where randomly selected households are asked to record all trips undertaken during a trial period. Providing the respondents are numerate and well sampled, the results of these surveys should be representative of the country as a whole. Here we use the UK National Travel Survey (NTS)~\cite{nts}, which has been carried out annually since 2002. Participants record all of their journeys for a week, and the trial periods are staggered throughout the year. The full data set includes the time, distance, purpose and mode of transport of nearly 2 million journeys.}

\par{More than 100,000 vehicles' usage can be extracted from the raw data. It is likely that most of these vehicles will be conventional, and the accuracy is limited by human error. However, as this type of data is cheap to gather, the sample size is much larger than any EV study. Similar surveys are conducted in other countries, such as the National Household Travel Survey in the US. The methods presented here could be applied to any data, where trip timings and distances recorded.}

\subsection{EV Trial Data}

\par{Data concerning EV use is currently scarce, however the results of some small-scale trials are available. This paper uses data from My Electric Avenue (MEA)~\cite{mea}, a UK trial which finished in 2016. During the 18 month trial period 213 Nissan Leafs were loaned out to households, with the caveat that all of their vehicle use and charging would be recorded and available for research purposes. The households were located in geographic clusters, and the trial was opt-in. This means that the behaviour captured is likely to represent early-adopters of EV technology and those living in the trial areas, but not the national as a whole.}

\par{In the vehicle usage data, the distance, time, and energy consumption of each journey are recorded. In the charging data, the time and state of charge (SOC) of the EV at the start and end of each charge are logged. Using the energy consumption of each journey and the fact that the vehicles had 24 kWh batteries, it is possible to infer the SOC of the vehicle at all times. For a more complete analysis of the data from this trial see \cite{Quiros-Tortos2018Mag}.}

\section{Cluster Analysis}\label{sec:clst}

\par{Clustering is used in this paper to reduce the dimensionality of vehicle usage as a model parameter. In this section we introduce the feature vector and algorithm used for clustering. Then analysis of the resulting clusters is presented, and the difference between the NTS and MEA vehicles' usage is quantified.}

\subsection{Feature Vector}

\par{Every point in the data to be clustered is defined by a feature vector, which is composed of a set of variables. Clustering groups points with similar feature vectors, so the choice of these variables dictates the action of the clustering algorithm.}

\par{Here, each vehicle-day is considered as a separate data point, meaning that a single vehicle from the dataset can belong to different clusters on consecutive days. This is consistent with the way that vehicles are actually used, e.g. a vehicle could be used to commute on one day, but not the next. The common practice of separating weekday and weekend behaviour is also adopted~\cite{weekdays}.}

\par{The feature vector needs to capture both the length and timing of all journeys that the vehicle carries out on that day. Here vehicles' normalised velocity at half-hour resolution are used as the feature vector. This means that there are 48 variables each describing a half-hour period, and a non-zero value indicates that the vehicle is used in that time interval. The average velocity profile is inferred using the distance and length of the journeys completed (effectively assuming a constant driving speed), then the profile is scaled such that the variables sum to 1. Normalising sacrifices the total distance travelled information, however vehicles travelling further are likely to be used for longer, so this information is still captured. Normalising is a common choice in profile clustering, as it tends to result in a more even distribution of points between clusters.}

\subsection{Algorithm}

\par{Here we use K-means clustering, a simple algorithm which selects the clusters which minimize the inter-cluster variance. The K clusters are each defined by a centroid, given by:
\begin{align}
\mathbf{y}^{(c')} = \frac{1}{N_{c'}}\sum_i^{N_{c'}} \mathbf{x}_i^{(c')},
\end{align}
where $\mathbf{y}^{(c')}$ is the centroid of cluster $c'$,  $\mathbf{x}_i^{(c')}$ is the feature vector of the $i$th point belonging to cluster $c'$, and $N_{c'}$ are the total number of points in that cluster. Each point is assigned to the cluster whose centroid is closest to it, as measured by Euclidean distance. The algorithm finds the cluster centroids which minimize the inter-cluster variance (the spread of points within a cluster). This is probably the most commonly used clustering method, largely due to its computationally simplicity. More complex methods are available for time series clustering (e.g. \cite{kshape}), however K-means produced useful results, and the size of the NTS data made computational cost paramount in this exercise.}

\par
\begin{figure}
\centering
\includegraphics[width=8.5cm]{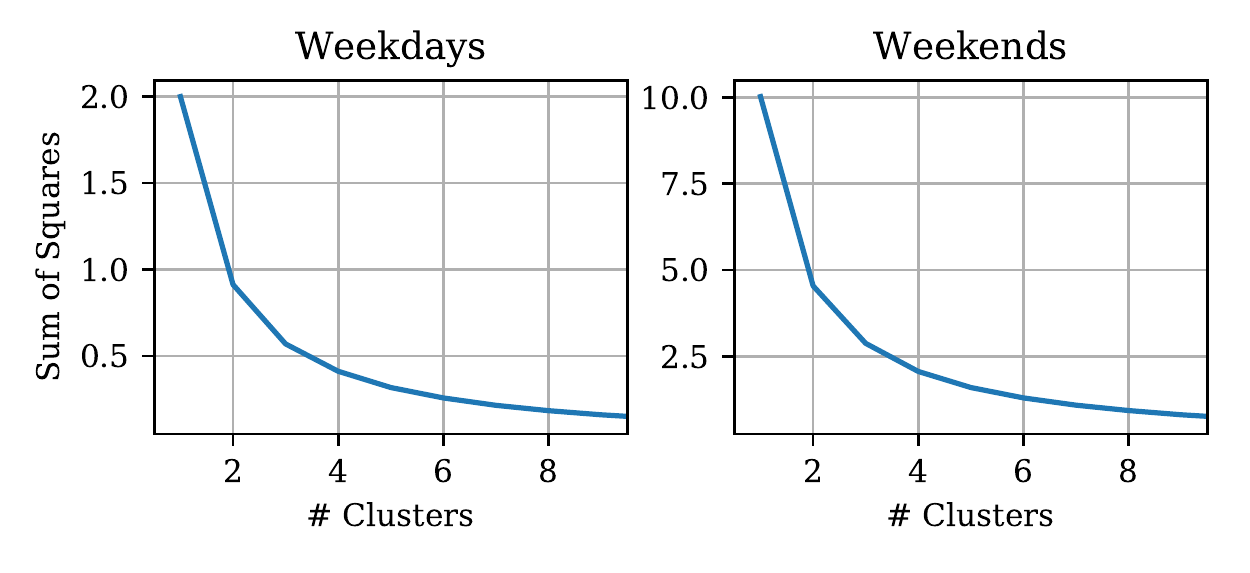}
\caption{The variation of sum of squares with number of clusters for both the weekday and weekend datasets.}
\label{fig:elbow}
\end{figure}

{One of the downsides of the K-means algorithm is that the number of clusters, K, needs to be defined. Various metrics have been proposed to do this, and here the elbow method is followed (e.g.\cite{elbow}). This method dictates that K is found by plotting the variation of \textit{sum of squares} with number of clusters. Sum of squares is defined as:
\begin{align}
SoS = \sum_i^N \left\Vert \mathbf{x}_i^{(c)} - \mathbf{y}^{(c)} \right\Vert^2,
\end{align}
where $N$ is the total number of data points across all of the clusters. This is a measure of inter-cluster variance, and will necessarily decrease as K is increased. K is then chosen at the \textit{elbow} (or the corner point) of this curve, where the reduction in variance achieved by an additional cluster is no longer significant. Fig. \ref{fig:elbow} shows the curve for both the weekday and weekend data, in both cases K=3 was selected. Implicitly there is an extra cluster containing vehicles which are not used in that day; these all have zero feature vectors and are removed before the clustering process. It is worth noting that the variance is higher for the weekend data than the weekday, even though there are fewer weekend days. This shows that weekend driving behaviour is more variable, meaning it is likely harder to predict and model.}

\subsection{Resulting Clusters} \label{sec:rc}

\par
\begin{figure}
\subfloat[Weekday clusters]{%
  \includegraphics[clip,width=\columnwidth]{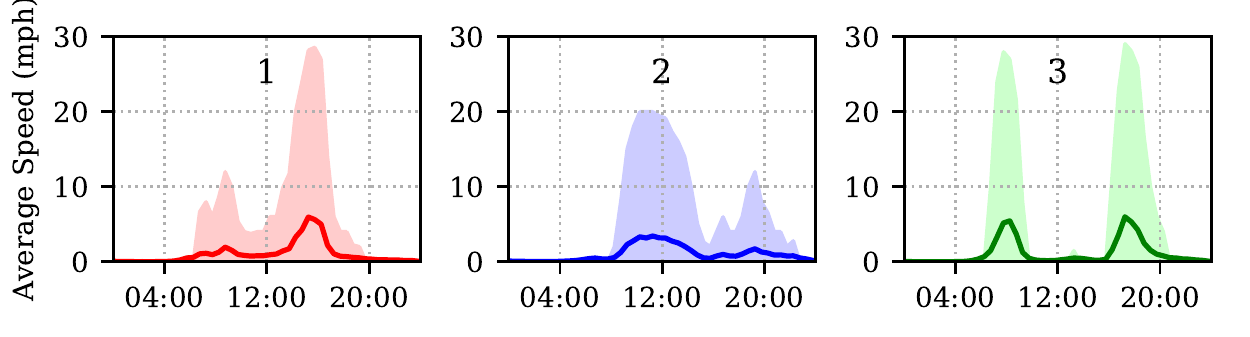}%
}

\subfloat[Weekend clusters]{%
  \includegraphics[clip,width=\columnwidth]{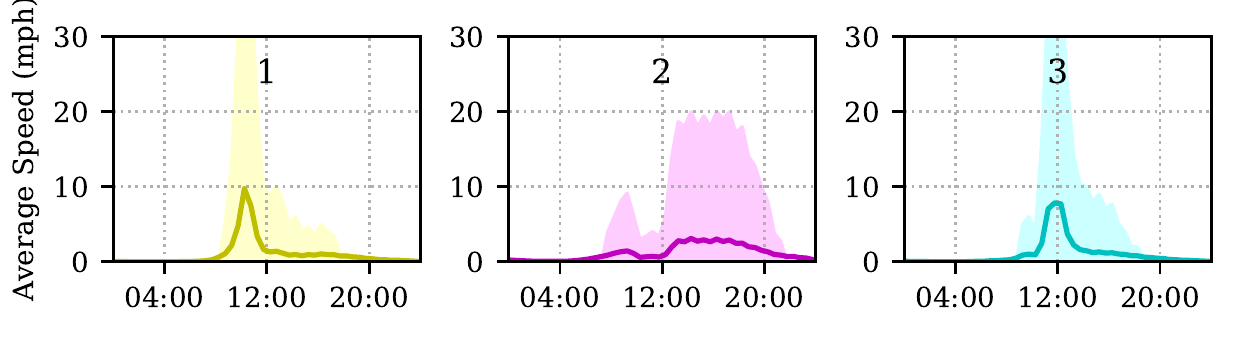}%
}
\caption{The average speed profile of the vehicles in each cluster. The lines show the mean values, and the shaded areas cover the 90\% confidence interval. There is no significance to the ordering of the clusters.}
\label{fig:clusters}
\end{figure}

\begin{figure}
\centering
\includegraphics[width=\columnwidth]{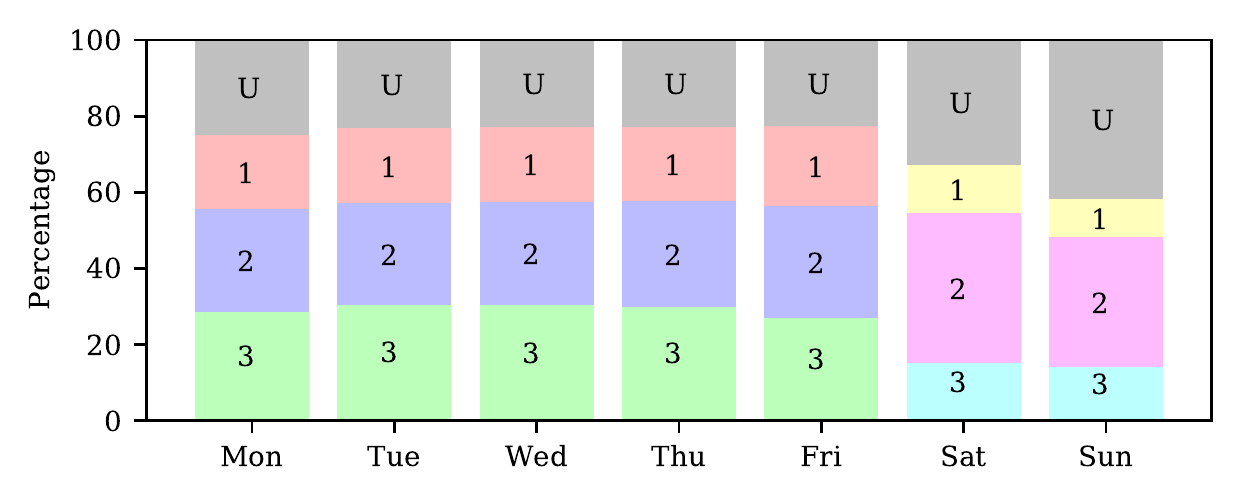}
\caption{The percentage of each cluster occurring on each weekday. The colours correspond to those in Fig. \ref{fig:clusters} except grey which indicates unuse.}
\label{fig:com}
\end{figure}

{Fig. \ref{fig:clusters} shows the average speed profile of vehicles from each cluster. It is worth noting that these are not the same as the cluster centroids, because the points are not normalised before averaging. For the weekdays: cluster 3 follows a typical commuting pattern, 1 is dominated by evening use, and 2 by morning use. In the weekends: clusters 1 and 3 suggest a single short journey at different times, while 2 shows more distributed use throughout the day. Fig. \ref{fig:com} shows the weekly composition of clusters, where the colours correspond to those in Fig. \ref{fig:clusters}. It can be seen that vehicle use is fairly consistent across the weekdays, although commuting is slightly less common on Mondays and Fridays. Overall vehicle usage is lower at the weekends, and lowest on Sunday.}

%\begin{table}[]
%\centering
%\caption{The average distance travelled by each cluster (miles)}
%\label{t:uk}
%\begin{tabular}{cccc}
%\toprule
%        & \multicolumn{3}{c}{\bf{Cluster}} \\
%        & \bf{1}       & \bf{2}       & \bf{3}       \\ \midrule 
%Weekday &     25.42    &    25.41     &    27.17     \\ 
%Weekend &     23.87    &    28.10     &      25.15  \\ \bottomrule
%\end{tabular}
%\end{table}

\par{When considering the variability in vehicle use, it is important to distinguish the variation between vehicles, from that in a single vehicle's behaviour. If the latter is small, then predicting charging at distribution system level becomes much simpler, as it is likely that the EV charging demand on a given feeder will be similar every day. Here the consistency of vehicle use is quantified with the cluster transition probabilities for consecutive days. Table \ref{t:trans} shows the probability that a vehicle belonging to one cluster will belong to another on the next day, U indicates that the vehicle was unused on that day. It can be seen that, although the most likely estimate is always that a vehicle will belong to the same cluster, many of the vehicles are used in different ways on consecutive days. Cluster 3 (the commuting cluster) exhibited the most consistent behaviour -- with 62\% of vehicles also commuting the next day. This implies that variation in vehicle use must be considered even at low aggregation levels.}

\begin{table}[]
\centering
\caption{The transition probabilities for vehicles on weekdays.}
\label{t:trans}
\begin{tabular}{cccccc}
\toprule
    &  & \multicolumn{4}{c}{Next day cluster} \\
      &  & \bf{1}       & \bf{2}       & \bf{3}     & \bf{U}       \\ \midrule 
& \bf{1}   &     40.2\%    &    27.6\%     &    17.6\%     &    14.6\%     \\ 
Current& \bf{2}   &     19.4\%    &    43.5\%     &    16.9\%     &    20.2\%     \\ 
cluster& \bf{3}   &     12.7\%    &    16.1\%     &    62.0\%     &    9.2\%     \\ 
& \bf{U}   &     14.9\%    &    27.6\%    &    14.1\%     &    43.4\%    \\ \bottomrule
\end{tabular}
\end{table}

\subsection{Comparison to Existing Electric Fleet}\label{sec:comp}

\par
\begin{figure}
\centering
\includegraphics[width=\columnwidth]{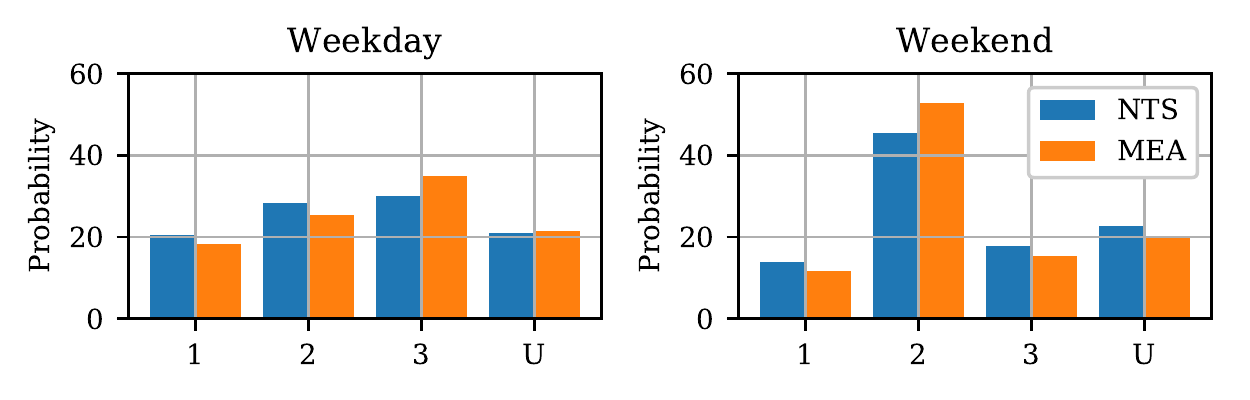}
\caption{A comparison of the cluster composition of the NTS and MEA data.}
\label{fig:MEAvNTS}
\end{figure}

{MEA provides the best available evidence for EV user residential charging behaviour in the UK. However, the vehicle use exhibited represents a biased set of drivers -- 67.3\% of participants were male, and 41\% were within the 40-49 age bracket. Quantifying this bias allows the likely error from extrapolating this trial data to represent a large fleet of vehicles to be predicted. Here this is achieved by creating equivalent feature vectors from the MEA data and classifying points according to the clusters defined in Section \ref{sec:rc}. By comparing the cluster composition of the datasets, modes of vehicle use which are over represented in the trial data can be identified. Fig. \ref{fig:MEAvNTS} shows the distribution of clusters for both datasets, and Table  \ref{t:uk} shows the average daily distance travelled by vehicles in each cluster.}

\begin{table}[]
\centering
\caption{The average distance travelled by each cluster (miles)}
\label{t:uk}
\begin{tabular}{ccccccc}
\toprule
        & \multicolumn{3}{c}{\bf{NTS Cluster}} & \multicolumn{3}{c}{\bf{MEA Cluster}} \\
        & \bf{1}       & \bf{2}       & \bf{3}       & \bf{1}       & \bf{2}       & \bf{3}       \\ \midrule 
Weekday &     25.42    &    25.41     &    27.17 &     28.83    &    29.33     &    29.70     \\ 
Weekend &     23.87    &    28.10     &      25.15  &     22.49    &    26.47     &      24.61  \\ \bottomrule
\end{tabular}
\end{table}

\par{The cluster composition is broadly similar, although there is a slight bias in the MEA data towards weekday commuters. However, distance travelled varies more significantly -- all weekday clusters travel further than average and all weekend clusters travel shorter distances. Overall the average MEA driver travels 12\% further than the average NTS driver on a weekday. Therefore, using the MEA data to directly forecast future charging is likely to produce overestimates.}

\section{Modelling Charging}\label{sec:cm}

%\par{This section describes the stochastic model which is used to map vehicle usage to a charging profile. First, we will define two different modes of charging observed in the MEA dataset. Then, the random variables considered are defined and the resulting probability distributions analysed. Finally, the modelling framework is described.}

\par{The most prevalent assumption in the literature is that EV charging begins immediately after the completion of the final journey~\cite{Huang2010,Darabi2011,Yan2017,Barghi-Nia2015,Pashajavid2012,Klayklueng2015,Ahmadian2015}. However, only 41\% of charges recorded in the MEA data begun within 10 minutes of finishing their final journey. Furthermore, only 70\% of charging events were within 10 minutes of the completion of any journey. This complicates the prediction problem, because there are a small number of journey end times, and a comparatively large number of times from which the other 30\% of charges will occur. Therefore, we propose considering these types of charging as distinct, and they will be hereafter referred to as \textit{after journey} and \textit{independent} charges.}

\par{The charging captured in MEA will reflect the usage from the trial, however in the long term the NTS will likely be more representative of EV driving patterns. The proposed model is built on the premise that people who use their vehicles similarly, will exhibit similar charging behaviour. This allows us to parameterize the model with the MEA data, while still applying it to the NTS data. This means that the bias created by the trial participants should not be present in the results.}

\par
\begin{figure}
\subfloat[Weekday]{%
  \includegraphics[clip,width=\columnwidth]{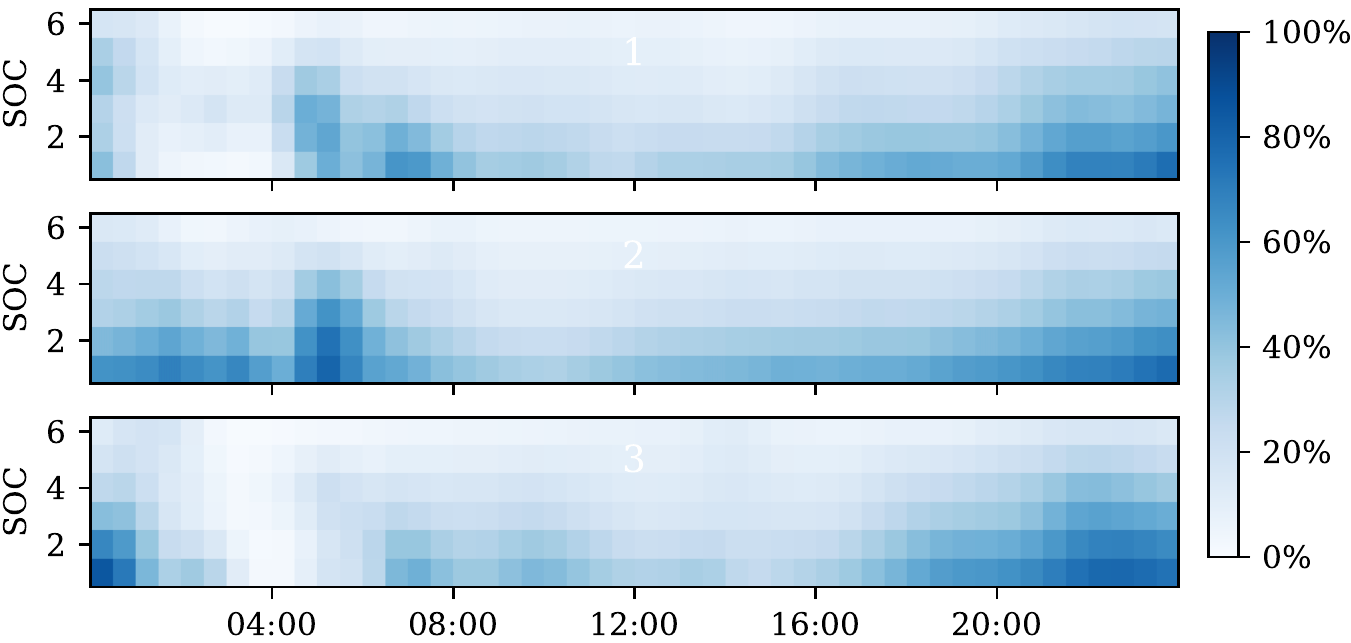}%
}

\subfloat[Weekend]{%
  \includegraphics[clip,width=\columnwidth]{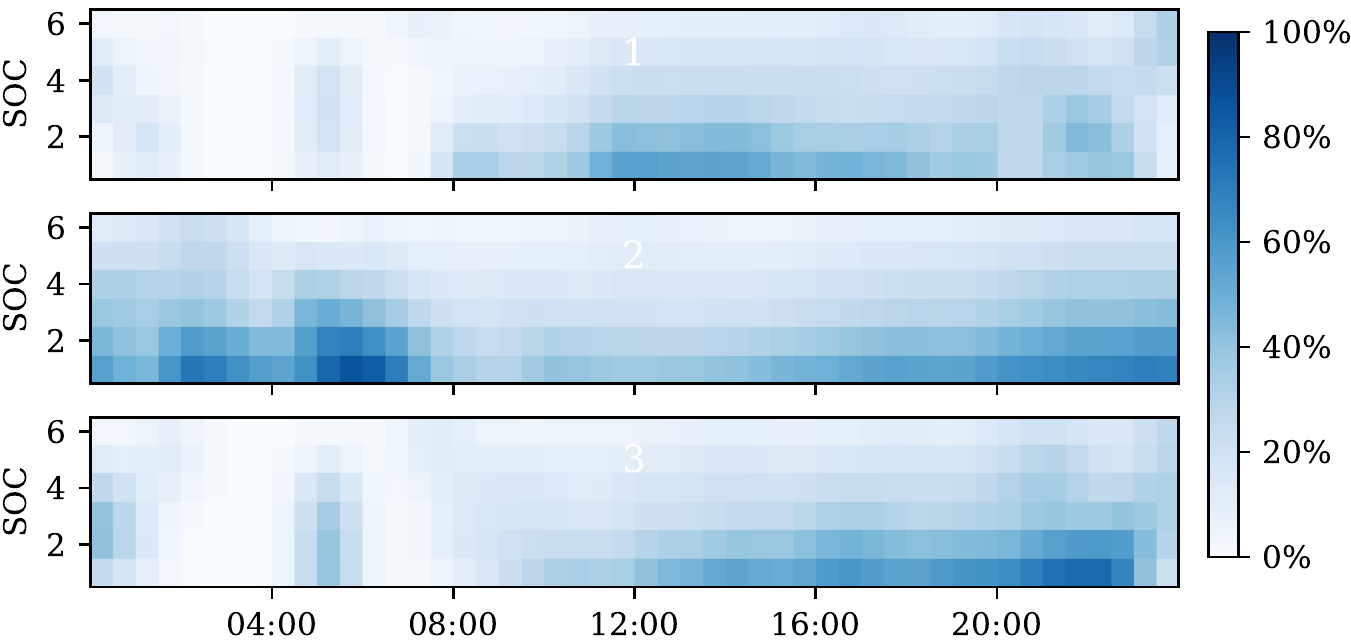}%
}
\caption{The \% probability that a charge will follow the completion of a journey, as a function of both time and SOC, for each vehicle use cluster.}
\label{fig:hm}
\end{figure}

{In the proposed model, the variables considered to influence charging decision are: the vehicle's SOC, the time, and the usage cluster that the vehicle belongs to. SOC is discretised into 6 states, and time is discretised into 48 half hour states. Formally, we define the following random variables:
\begin{align*}
c_j \in\mathbb{Z}_{2},\quad c_i \in\mathbb{Z}_{2},\quad d \in\mathbb{Z}_{2},\\
k \in\mathbb{Z}_{3},\quad t \in\mathbb{Z}_{48}, \quad s \in\mathbb{Z}_{6},
\end{align*}
where $\mathbb{Z}_{x}$ denotes the integer set from 1 to $x$, $c_j$ is the binary variable determining whether an after journey charge begins, $c_i$ the binary variable determining whether an independent charge begins, $d$ states whether it is a weekday or weekend, $k$ is the cluster the vehicle belongs to that day, $t$ is the time, and $s$ is the SOC. Now instead of considering only the probability that a charge will occur, we must consider the \textit{joint distribution} of all variables. Every possible scenario is described by a combination of these variables, meaning that: 
\begin{align}
\sum_{c_j,c_i,d,k,t,s} P(c_j, c_i, d, k, t, s) = 1,
\end{align}
where $P$ is the probability distribution function. The prediction problem becomes calculating the \textit{posterior} probability that a charge begins, given the known values for the other variables. This is written as:
\begin{align} \label{eq:ajpost}
P(c_j = \text{True}\mid c_i, d, t, k, s),
\end{align}
where $\mid x$ implies that the value of $x$ is known. From the definitions of $c_i$ and $c_j$ it can be seen that it is impossible for both variables to be true simultaneously, as they are describing the same phenomena under different circumstances. Therefore we can exclude $c_i$ from (\ref{eq:ajpost}) because if $c_j$ is true then $c_i$ can only be false. The expression therefore reduces to:
\begin{align}  \label{eq:cj}
P(c_j = \text{True}\mid d, t, k, s),
\end{align}
which is defined over $2\times48\times3\times6=1728$ possible scenarios. This discrete distribution can be populated using the observed charging events from the MEA data. For each $(d,t,k,s)$ (\ref{eq:cj}) is approximated as the percentage of instances of those variables which resulted in a charge. Note that only times when a journey had just ended are considered. A Gaussian filter was used to smooth the distributions in order to compensate for areas in state space where data was scarce. Fig. \ref{fig:hm} illustrates (\ref{eq:cj}) using heatmaps, one for each possible $k$ and $d$ combination, with $t$ on the horizontal axis and $s$ on the vertical.}

\par{The fact that the distributions vary significantly with $k$ supports its incorporation as a parameter; if EVs' charging was independent of usage cluster the three heatmaps would be identical. The peaks occur at low values of SOC (as expected), in both the evening and early morning. Note that this does not mean that all vehicles are likely to charge in the early hours, just that those completing journeys at this time are.}

\par
\begin{figure}
\centering
\includegraphics[width=\columnwidth]{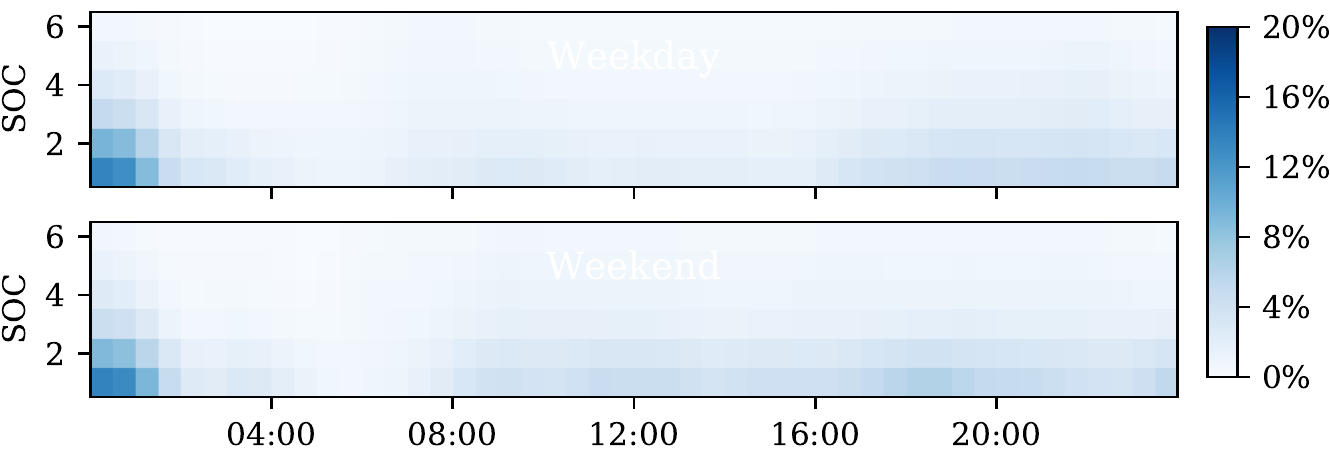}
\caption{The \% probability that a charge will start independant of a journey, as a function of both time and SOC, for each vehicle use cluster.}
\label{fig:rand_hm}
\end{figure}

{For independent charging it was found that the vehicle usage had a negligible effect on whether or not a charge was started. In fact, often these events occurred on days where there was no vehicle use -- and as a result no value of $k$. Therefore it was assumed that $c_i$ was independent of $k$, such that the posterior distribution to be estimated becomes:
\begin{align}  \label{eq:ci}
P(c_i = \text{True}\mid t, k, s),
\end{align}
Fig. \ref{fig:rand_hm} illustrates this distribution. In this case there is not significant difference between weekend and weekdays, suggesting that $d$ could also be excluded from (\ref{eq:ci}). However, as minor differences are observed in the early evening (which is the time of greatest interest) the variable was kept in this analysis. Here the distribution peak occurs shortly after midnight, and it is suggested that this is the results of timers set to coincides with the start of economy7 cheaper pricing.}

\par
\begin{figure}
\centering
\includegraphics[width=0.7\columnwidth]{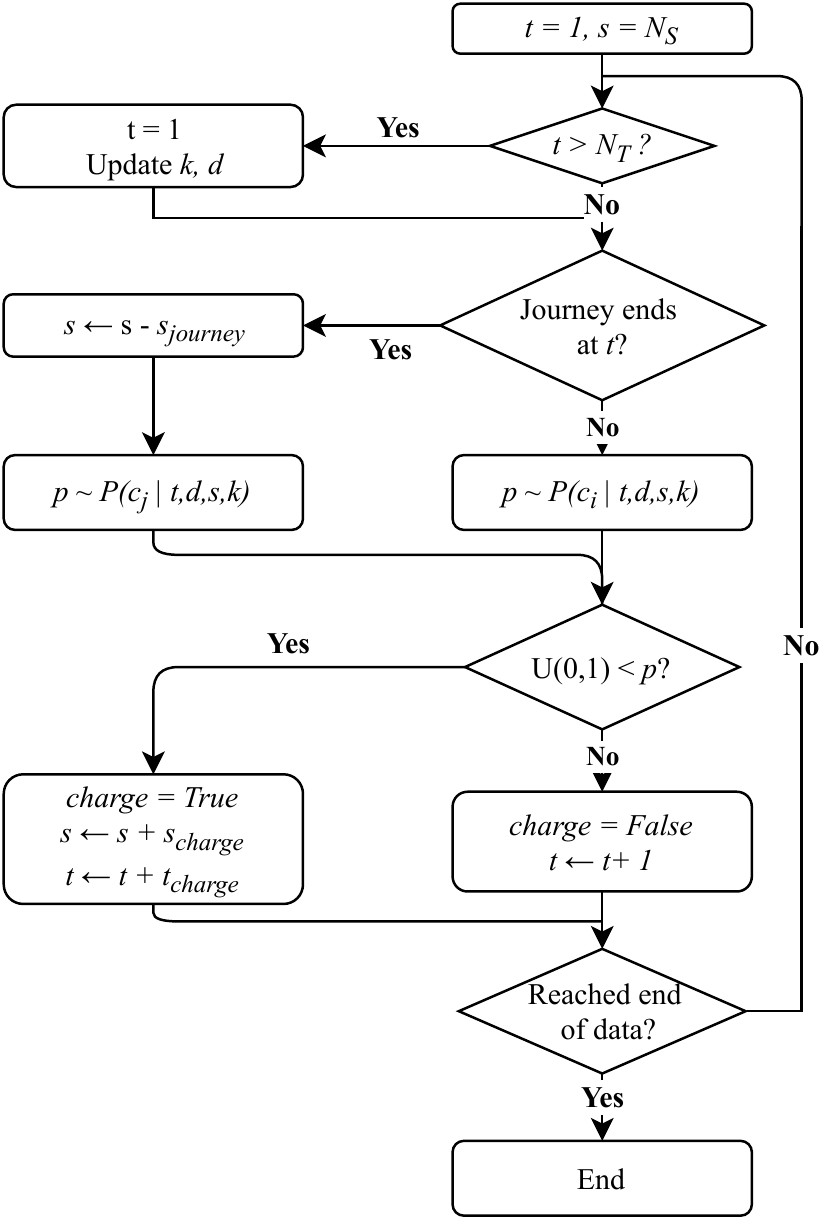}
\caption{A flow chart describing the charging model simulation process.}
\label{f:flowchart}
\end{figure}

{These distributions can then be applied to the NTS data, as $(d,t,k)$ are known and $s$ can be estimated by assuming a battery capacity and a fixed rate of energy consumption per mile. This allows a Montecarlo simulation to be set up, which is described by the flow chart in Fig. \ref{f:flowchart}. For each vehicle $(k,d,s,t)$ are initialised, then for incremental values of $t$:
\begin{itemize}
\item If a journey ends at t, sample $P(c_j = \text{True}\mid d, t, k, s)$  and reduce $s$ as necessary.
\item Otherwise, sample $P(c_i = \text{True}\mid t, k, s)$.
\item Sample the uniform distribution $U(0,1)$ and if it is less than the sampled number, begin charging.
\item Charging ends either when the battery is full, or the vehicle is next used. Update $s,t$ as necessary.
\end{itemize}}
%In this case $t$ is at 1 min resolution, so the values in Figure \ref{fig:rand_hm} are all scaled by $\frac{1}{30}$.}

%\par{If the vehicle completes a journey at $t$ then the probability that an after journey charge begins is found, given $(s,d,k,t)$, if not the probability that an independent charge begins is found, given $s,d,t$. Next we generate a random number, and if it is less than the probability calculated it is assumed that the EV plugs in. Charging ends either when the battery is full, or the vehicle is next used -- whichever occurs first. The process is repeated until the end of the usage data is reached, ensuring that when $t$ reaches midnight the values for $d$ and $k$ are updated.}

\par{Stepping through the data once will result in a single estimate of charging. Stochasticity is captured by repeating the simulation, resulting in a distribution of predicted charging. Variation in both charging and vehicle use can be incorporated by running further Montecarlo simulations where the input vehicles are randomly sampled from the travel survey..}

\section{Model Validation}\label{sec:val}

\par{The accuracy of the model proposed in Section \ref{sec:cm} can be quantified by predicting the charging of the MEA vehicles from their usage data. As the MEA vehicles were used to create the distribution, each vehicle was removed from the training data while its charging was predicted. For an individual vehicle-day the model predictions will vary significantly every time it is run. However, when considering the prediction of the whole dataset the law of large numbers says the variance should become very small. Therefore, an overall PDF of predicted charging start times can be produced by running the model a small number of times over the entire dataset.}

\par
\begin{figure}
\centering
\includegraphics[width=\columnwidth]{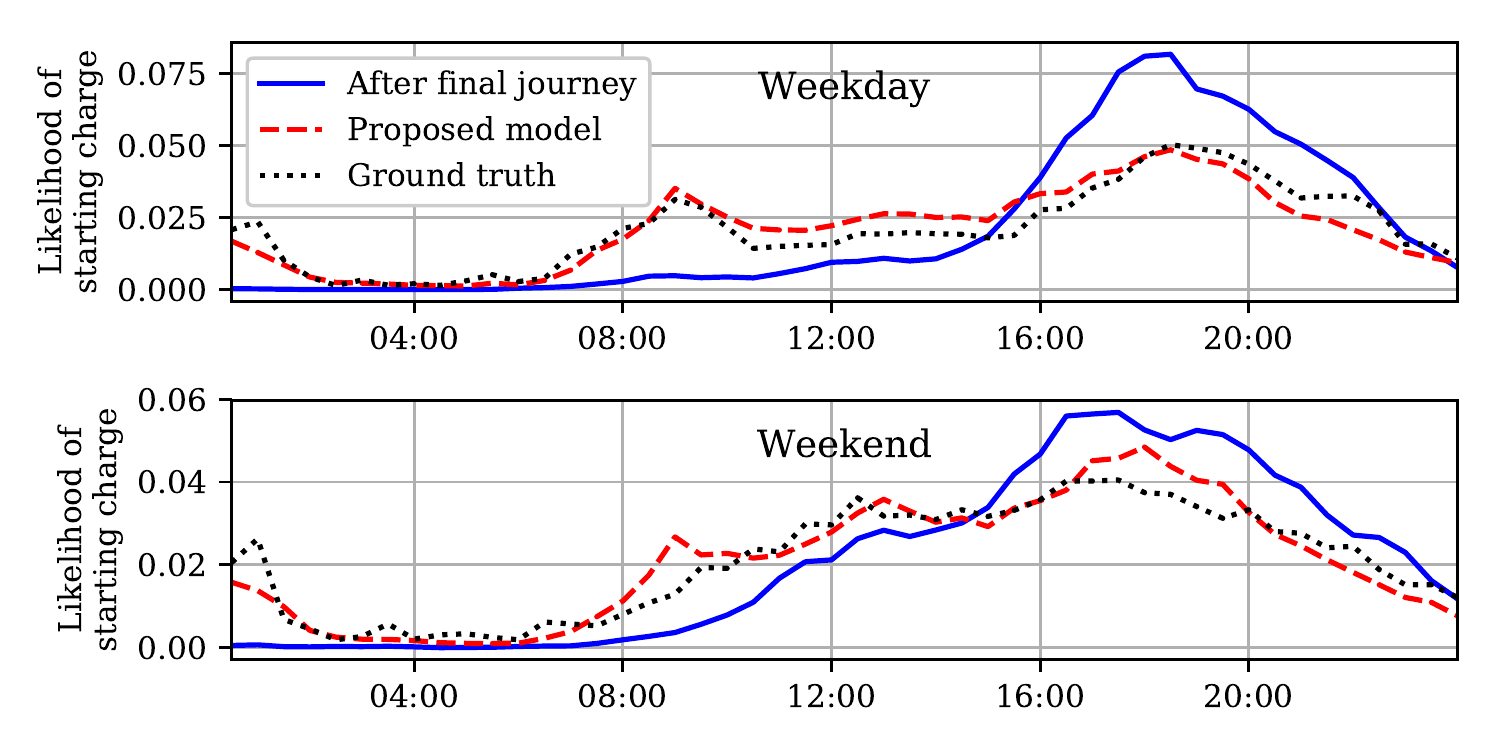}
\caption{The likelihood of starting charging predicted from the MEA usage assuming under both models. The true MEA charging likelihood is also shown.}
\label{f:error}
\end{figure}

{For charging after final journey there is no stochasticity in the model, so the PDF of starting charging will equal the PDF of final journey end times. Fig. \ref{f:error} shows both estimated PDFs, compared to the distribution of times when charging is observed to have started. It can be seen that assuming charging starts immediately after the last journey biases predictions between 16:00 and 23:00 by up to 70\%. This is especially problematic for peak demand prediction, as the existing peak occurs within this window. The new charging model achieves within 25\% error across all times. Weekday observations are fit with greater accuracy, which is unsurprising given the larger variability in weekend vehicle use. It could be argued that, as the weekday vehicle use is higher, these days are of greater concern.}

\par{Rather than the start of charging, network operators are more concerned with predicting the power demand profile of vehicles. This is effectively obtained via a convolution of the start of charging PDF, so the error will be lower for both models. This is confirmed in Table \ref{t:errors} which shows the mean absolute percentage error for both PDFs. It is worth noting that these metrics only show the expected error; the error on individual predictions will vary case-by-case.}

\begin{table}[]
\centering
\caption{Mean absolute percentage error in weekday distributions.}
\label{t:errors}
\begin{tabular}{rcc}
\toprule
& \bf{Starting charging} & \bf{Power demand} \\ \midrule 
Section \ref{sec:cm} model & 20\% & 18\% \\
After final journey & 58\% & 53\%   \\ \bottomrule
\end{tabular}
\end{table}

\section{Case Studies}\label{sec:res}

\par{In this section, example uses of the proposed model are demonstrated. First, we consider prediction of the aggregated charging of 50 vehicles in a single region. Second, we investigate the geographic variation of EV charging in the UK.}

\subsection{Residential network study}

\par{Here we consider the aggregated charging of 50 households' vehicles. This is representative of charging in a LV distribution network, where 100\% of vehicles are electric. Simulations of this kind are important, because we need to understand how diversity between vehicles is likely to manifest at low levels of aggregation. Likely, if 50 vehicles charged simultaneously on a single feeder then network limits would be violated. However there are existing appliances (e.g. kettles or showers) which would cause overload if all households used simultaneously; in reality natural diversity between users renders this situation extremely unlikely. As EV adoption increases, accurately modelling the diversity of EV charging will be crucial in predicting the peak demand.}

\par{For this case study, vehicle data was taken from NTS households in North Lincolnshire (a county in the North East of England) on a Wednesday. These parameters were chosen to remove geographic and weekday variations in vehicle use. It was assumed that chargers were rated at 3.5 kW and had an efficiency of 90\%. Montecarlo simulations were constructed to estimate the average and variance of the predicted charging profile. Two simulations were carried out, one considering only variation in charging, and one considering both variation in vehicle use and charging. In the first, a single set of 50 vehicles was chosen from the data, and in the second, the 50 vehicles were allowed to varied between runs of the Montecarlo simulation.}

\begin{figure}
\centering
\includegraphics[width=\columnwidth]{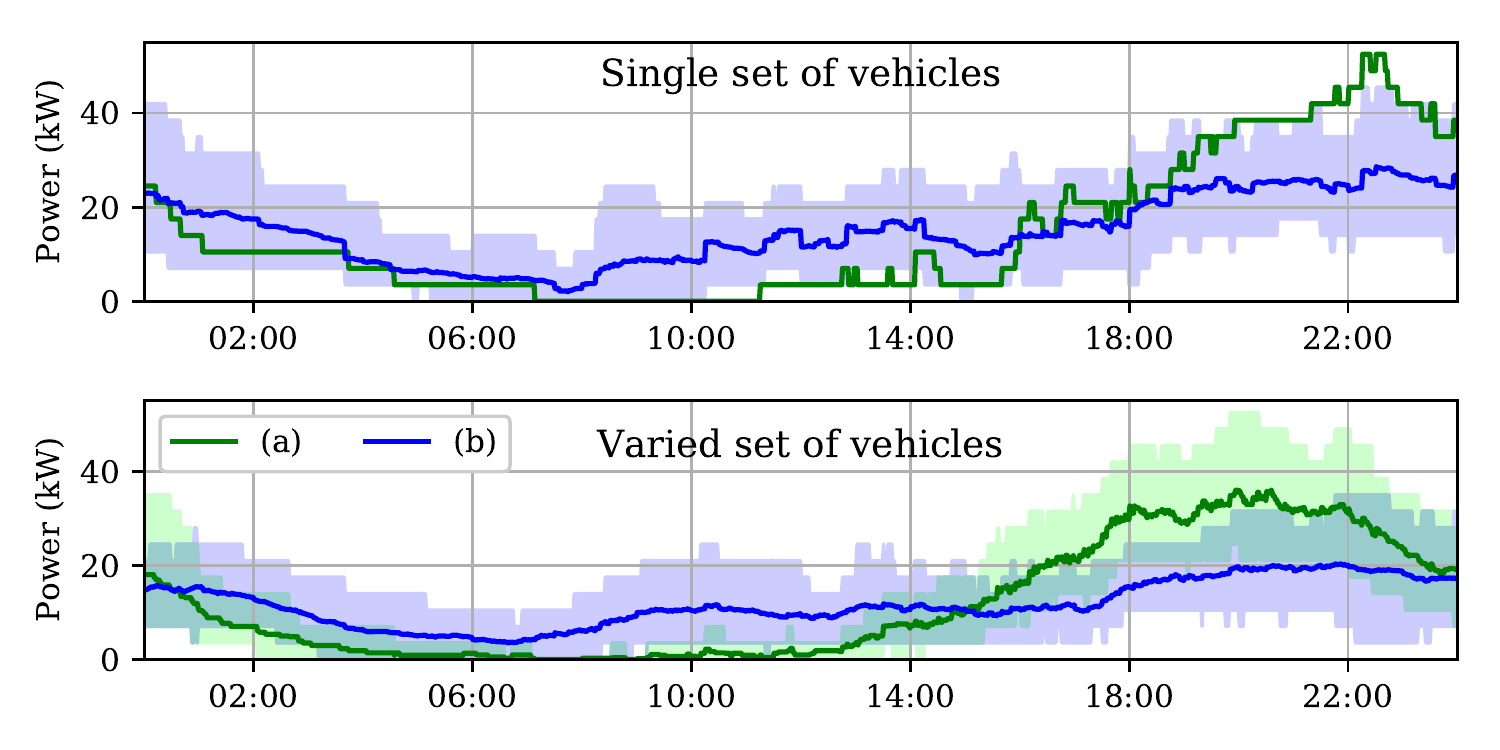}
\caption{Aggregated charging of 50 households' vehicles under both: (a) the assumption that charging always begins after a vehicle's final journey, and (b) the proposed model. The top plot shows the result for one set of vehicles, and the bottom takes into account variation is vehicles as well as charging. The shaded area covers the 90\% confidence interval of simulation results.}
\label{fig:mc_power}
\end{figure}

\par{The simulation results are shown in Fig. \ref{fig:mc_power}, for both the model in Section \ref{sec:cm} and the assumption that vehicles charge after their final journey. In the single set simulation there is no variation in the latter as the model is deterministic. Whereas, when the set of vehicles is varied, stochasticity is introduced via the vehicle use. In both simulations the peak demand predicted under the new model is lower than assuming charging after final journey. This is significant because it suggests that existing predictions of the impact of EV charging on distribution networks are overestimates. It can also be seen that, at this level of aggregation, there is a large variation in charging power -- even when there is no variation in vehicle use. This suggests that uncontrolled charging of EVs will add significant uncertainty, as well as magnitude, to residential networks' load.}

\subsection{UK geographic variation}

\par{Due to the abundance of travel survey data available, it is possible to compare the likely impact of EV charging in different areas, assuming no change in driving patterns. This section uses the NTS data to estimate the regional variation EV charging impact of LV distribution networks.}

\par{In the UK, networks are designed to tolerate a certain after diversity maximum demand (ADMD)~\cite{admd}. This is the peak demand at a 30-min resolution averaged over the number of households on a network. EV charging is likely to increase ADMD but some areas may be worse affected than others due to: larger travel distances, higher vehicle ownership, low variability between local vehicles, or a low existing peak load. A simulation was constructed estimating the increase in ADMD due to charging of a 100\% EV fleet in each local authority in the UK. This requires both the existing ADMD and the peak EV charging demand at 30 minute resolution to be predicted.}

\par{To quantify the charging demand, Montecarlo simulations were carried out for each local authority. In each run 50 households were randomly selected from the relevant NTS data, and their charging was predicted using the model described in Section \ref{sec:cm}. This process was repeated 200 times and the average charging profile was stored.}

\par{Existing ADMD depends on a number of factors, including: the energy efficiency of buildings, the number of residents per dwelling, the affluence of the area, and whether or not the homes are connected to the gas network. The UK government publishes the annual electricity consumption of households within each local authority, as well as the number of households on each tariff structure~\cite{elecLSOA}. In the UK there are two commonly available structures: a flat rate, and economy7 (where there are 7 hours at a lower rate). Elexon produce demand profiles representing the average flat-rate and economy7 user, on weekdays/weekends and in different seasons~\cite{elexonStd}. Here we estimate the existing power demand for each area by blending the profiles from the two tariff structures according to the percentage of homes on each meter type, and scaling according to the total energy consumption. The predicted charging is then superimposed onto this demand, and the percentage change in ADMD is calculated.}

\par
\begin{figure}
\centering
\includegraphics[width=0.87\columnwidth]{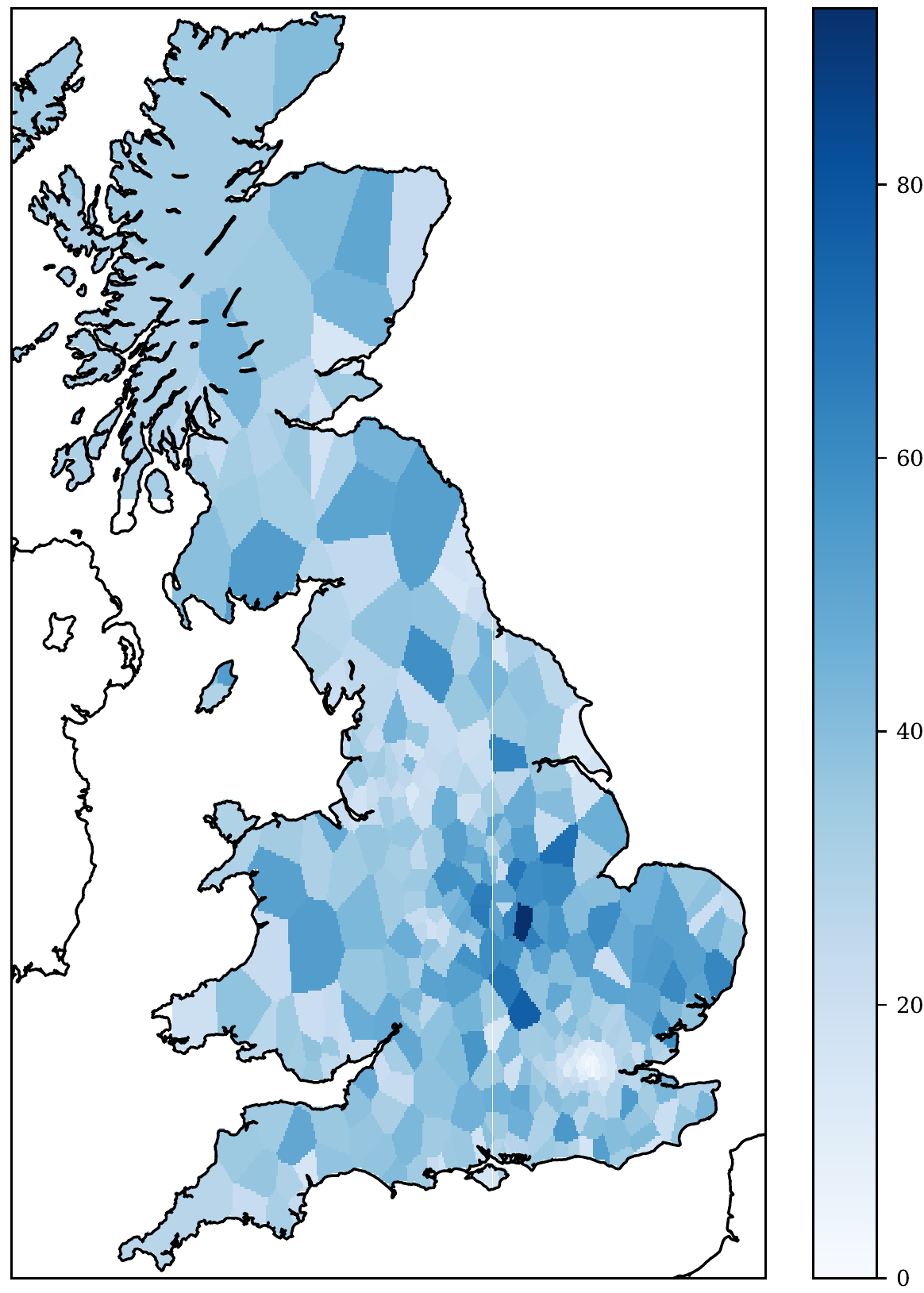}
\caption{The geographic variation in projected \% increase of winter LV network ADMD in the UK due to at home charging of a 100\% EV fleet.}
\label{f:uk}
\end{figure}

{Fig. \ref{f:uk} shows a map of the UK where each local authority is shaded to represent its predicted \% increase in ADMD due to EV charging. In London the increase was the smallest, likely because the public transport network is extensive, and congestion charges discourages private vehicle use. The worst percentage increase was seen in the Midlands, where vehicle use is high and existing household electricity demand is relatively low.}

\section{Conclusion} \label{sec:con}

\par{In this paper a stochastic model for EV charging was presented, which is parameterized by trial data, but applied to survey data. The model was based on conditional probability distributions, formulated from the trial data, and incorporates random variables for: vehicle usage, SOC, time, and type of day. K-means clustering was used to identify 3 distinct vehicle usage modes, and the cluster number was included as a model parameter -- allowing vehicle use to be incorporated as a single parameter. These variables can all be estimated from conventional vehicle data, so the probability of charging can be inferred for vehicles from travel survey data.}

%\par{In this paper we have developed a stochastic model for uncontrolled charging of EVs, which uses cluster analysis. K-means clustering was used to identify 3 distinct vehicle usage modes, and it was shown that users exhibited different charging behaviour between them. Charging was segregated according to whether it occurred directly after a journey or not. A conditional probability model was parameterized using data from a small EV trial, where charging was assumed dependant on: vehicle usage, SOC, time, and type of day.}

\par{The model correctly predicted 80\% of charges from the EV trial data within 10 minutes, whereas assuming charging occurred after completion of the final journey only achieved 42\%. The predicted peak demand of aggregated EVs' charging was 30\% lower using this model rather than the traditional assumption. It was also shown that the trial vehicles travelled further distances than average. This means that either applying simple charging assumptions to survey data, or extrapolating charging of trial data, is likely to overestimate charging demand. This suggests that the allowable penetration of EVs at the distribution level may be higher than initially estimated.}

\par{A case study was set up which investigated the regional variation of projected increase in ADMD in the UK under 100\% penetration of EVs. It was found that in London there would be only a marginal increase, while in the Midlands ADMD is likely to nearly double. This highlights the importance of accounting for regional variation, rather than using a single top-down model for vehicle charging for all regions.}

% or


\section*{Acknowledgment}
The authors would like to thank Jaguar Land Rover for their support in this work.

\ifCLASSOPTIONcaptionsoff
  \newpage
\fi

% number -- used to balance the columns on the last page
% adjust value as needed -- may need to be readjusted if


%\bibliographystyle{IEEEtran}
\bibliography{references.bib}{}
\bibliographystyle{IEEEtran}


%\end{IEEEbiography}


%\vfill


\end{document}